\documentclass[pra,aps,twocolumn]{revtex4}
\usepackage{amsfonts}
\usepackage{amsmath}
\usepackage{graphicx}
\usepackage{dcolumn}
\usepackage{bm}
\usepackage{epsfig}

\begin{document}
\title{Non-linear dynamics of double-cavity optical bistability of three-level ladder system}
\author{H. Aswath Babu and Harshawardhan Wanare}
\affiliation{Department of Physics, Indian Institute of Technology Kanpur, Kanpur 208016, India}
\date{\today}

\begin{abstract}
We present non-linear dynamical features of two-photon double-cavity optical bistability exhibited by a three level ladder system in the mean field limit. The system exhibits a hump like feature in the lower branch of the bistable 
response, wherein a new region of instability develops. The system displays a range of dynamical features varying from normal stable switching, periodic self-pulsing to a period-doubling route to chaos. The inclusion of two competing cooperative atom-field couplings leads to such rich nonlinear dynamical behavior. We  provide a domain map that clearly delineates the various regions of stability that will aid the realization of any desired dynamics.
We also present bifurcation diagram and the associated supporting evidence that clearly identifies the period-doubling route to chaos, which occurs at low input light levels. 
\end{abstract}

\maketitle
\section{Introduction}
\label{secintro}

Understanding instability is pivotal to fabricating practical devices~\cite{strogatz}. In recent
times chaos has been used in a variety of applications including generation of random numbers~\cite{chaos_ap1,chaos_ap2}, ultra-wide bandwidth~\cite{chaos_ap3}, optical communication scheme~\cite{chaos_ap4}, etc. Optical Bistability (OB)~\cite{gibbs,lugiato1984} has historically offered a platform for studying a  variety of nonlinear dynamical effects~\cite{ikeda1980,lugiato1981,joshi_modern}. In conventional OB the nonlinear dynamics arises due to the interplay of the atom-cavity coupling as well as the cavity characteristics such as the number of modes, cavity decay and detuning~\cite{bonifacio1976,bonifacio1978}. We describe a system wherein the nonlinear dynamics arises
due to the interplay of cooperative atom-cavity coupling at two different frequencies.  This results in a hitherto unseen nonlinear dynamical regime that arises in double-cavity two photon OB, where both the fields experience independent feedback. This new regime occurs at low input light levels, moreover the system also exhibits negative hysteresis bistable response~\cite{aswath_1}. We undertake a systematic study that enables a greater understanding of the phase space structure and thus allows us to selectively {\em steer} the system to exhibit stable self-pulsing, chaos or regular stable switching. Such control mechanisms offer the possibility of using these dynamics for communication technologies involving multiwavelength operation, metrology, and moreover this system offers 
a different paradigm to study the fundamental aspects of optical instabilities.   

Optical instability has been extensively studied in the last four decades. Ikeda et.al.~\cite{ikeda1980} have shown theoretically that periodic instabilities and chaotic behavior can occur in optical bistable systems with delayed feedback. Single mode instabilities ranging from gain based laser systems~\cite{haken1975} to passive two-level optical bistable systems have been investigated earlier~\cite{{orozco1989},{maize2006}}. Various studies have been undertaken related to the three-level atoms interacting with multiple fields leading to chaos: Savage et.al~\cite{savage1982} describe the possibility of tri- and quadra-stability as well as self-pulsing and chaos, Grangier et.al~\cite{grangier1992} have studied OB in the purely dispersive limit and shown the occurrence of chaos for large field intensities. Chaos has also been demonstrated in a three-level $\Lambda$ system with only the probe field
experiencing feedback and the coupling field detuning is used to drive the system to chaos~\cite{amitabh&min2005}. To the best our knowledge the regime we describe has not been reported earlier, we obtain chaos for moderate cooperative
parameters in the lower branch of the bistable response at low input intensity levels. These effects crucially rely on the interplay of two cooperative branches within the atom.  No approximations are made with regard to the nonlinearity of the active medium, the effects are quite robust and occur in a wide parameter regime. This model is ideal for investigation of optical chaos that arises just beyond the single mode limit, as we consider only two single modes associated with the two monochromatic fields coupling the two-photon transition. Moreover, as the system exhibits coordinated dynamics (such as self-pulsing) at two distinct optical frequencies, it can be gainfully utilized in optical communication technology. 

The organization of the paper is as follows: In section-~\ref{sectheo} we present the description of the theoretical model. We provide a detailed stability domain map as well as the details of the numerical modelling in section-~\ref{secnume} along with the various tools used in the analysis. In section-~\ref{secrana} we present the results  and  analysis of the nonlinear dynamical features including bifurcation diagrams, spectra etc.  We conclude in section-~\ref{secconc}.

\section{Theoretical Model}
\label{sectheo}
We consider two optical fields in two independent unidirectional ring cavities that share a region containing a collection of three-level ladder atomic media. The optical fields couple two adjacent transitions in the atom with the ground state $|3\rangle$, and intermediate state $|2\rangle$ and the excited state $|1\rangle$. The electric field at the atom 
\begin{equation}
E=E_1 e^{i\omega_1 t}+E_2 e^{i\omega_2 t}+c.c.,
\end{equation}
consists of two monochromatic fields of amplitude $E_1$ and $E_2$, at frequencies $\omega_1$ and $\omega_2$, coupling the transitions $|1\rangle\leftrightarrow |2\rangle$ and $|2\rangle\leftrightarrow |3\rangle$ respectively, identical to the system described in the companion paper~\cite{aswath_1}. The fields within the active medium are counterpropagating in order to minimize the effects of Doppler broadening in the ladder system. Our analysis is within the uniform field approximation also known as the mean field limit. 

\begin{figure}[thb]
\hspace*{-2.5cm}
\includegraphics[width=0.8\textwidth]{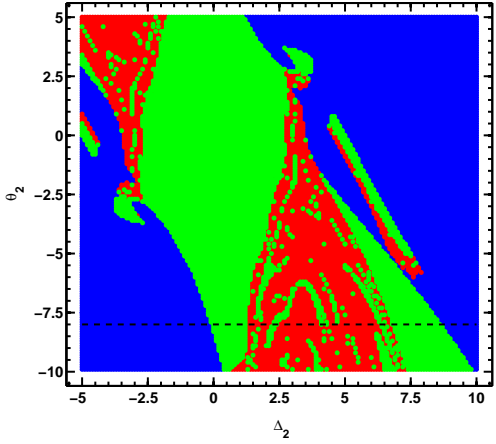}
\caption{(Color online)The stability domain map plotted between $\Delta_2$ and $\theta_2$, indicating the 
stable fixed point region(blue) and the self-pulsing region(green) and the chaotic region(red), for the parameters 
$\gamma_i=1,\kappa_i=1,\Delta_1=0,\theta_1=0,C_i=200,|y_1|=23,|y_2|=40$, for $i = 1,2$. 
The dashed line ($\theta_2 = -8$) indicates the particular parameter variation we consider in Fig.~\ref{undfigroute} to study the bifurcation.}  
\label{domain-map}
\end{figure}

We consider the usual boundary conditions imposed independently on the two fields,  due to the cavity feedback~\cite{lugiato1984} 

\begin{eqnarray}
\nonumber
E^{out}_{i}(t)&=&\sqrt{T_{i}} E_{i}(L,t) \\ 
E_{i}(0,t)&=&\sqrt{T_{i}} E^{in}_{i} +R_i e^{-i\delta_{i}} E_{i}(L,t-\Delta t)
~\label{bc}
\end{eqnarray}
for the field $\omega_i$. The subscript $i = 1, 2$, in this paper, refers to the two fields at $\omega_1$ and $\omega_2$, respectively. Here, $T_i$ and $R_i$ are the transmission and reflection coefficients associated with the two independent cavities, and $L$  is the length of the active medium. The Eqs.~\ref{bc} relate the input-output fields of the cavity. For simplicity we have assumed that the time taken by both fields outside the active medium is identical and is represented as $\Delta t$. The cavity detunings are $\delta_i=(\omega^c_i - \omega_i){\cal{L}}_i/c$, where $\omega^{c}_i$ is the nearest resonant cavity frequency close to the field frequency $\omega_i$, and ${\cal{L}}_i$ is the total length of the corresponding cavities. The wave equation under the slowly varying envelope approximation is 

\begin{eqnarray}
\frac{\partial E_i}{\partial  t} + c \frac{\partial E_i}{\partial  z}  = i 2 \pi \omega_i P(\omega_i)
~\label{waveeqn}
\end{eqnarray}

where $P(\omega_i)$ is the macroscopic atomic polarization, which is obtained from the density matrix of the atomic system described below, given as $P(\omega_1)=n d_{12}\rho_{12}$ and $P(\omega_2)=n d_{23}\rho_{23}$ where $n$ is the number density of atoms, $d_{lm}$ and $\rho_{lm}$ are the dipole matrix elements and the atomic density matrix elements comparable to the $|l\rangle \leftrightarrow |m\rangle $ transition, and are described in detail below. 

\begin{figure}[thb]
\centering
\includegraphics[width=0.5\textwidth]{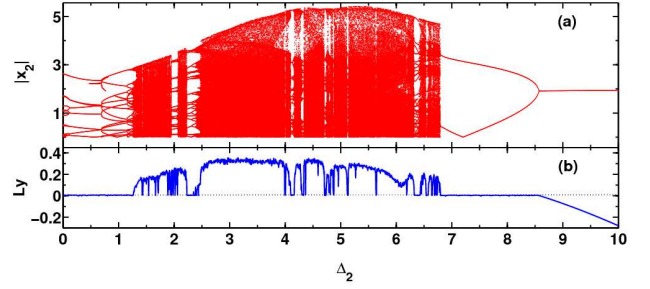}
\caption{(Color online)(a) The bifurcation diagram indicating the out put field $|x_2|$ and (b) The largest Lyapunov exponent for th same is plotted with respect to $\Delta_2$ for $\theta_2=-8$. The parameter values are same as those specified in Fig.~\ref{domain-map}.}
\label{undfigroute}
\end{figure}

We undertake the analysis for the two fields along the lines of Ref.~\cite{lugiato1984}, and Eq.~(\ref{waveeqn}) can be reformulated using the above boundary conditions Eq.~(\ref{bc}) as 

\begin{eqnarray}
\frac{\partial E_i}{\partial  t'} + c \frac{L}{\cal{L}}_i\frac{\partial E_i}{\partial  z} 
& =& \kappa_i \left[-E_i(1+i \theta_i)+\frac{ E_i^{in}}{\sqrt{T_i}}\right] \nonumber \\
& + & \frac{i 2 \pi L\, \omega_i}{{\cal{L}}_i}\,P(\omega_i)
\end{eqnarray} 

where $t'=t+\Delta t \frac{z}{L}$, the cavity decay  $ \kappa_{i} = c T_{i}/{\cal{L}}_i$, the cavity detuning is scaled with $\kappa_{i}$ and is given as $\theta_i=\delta_i/\kappa_i$. The absorption coefficients $\alpha_i$ along the two transitions are 

\begin{equation}
\nonumber
\alpha_1=\frac{2\pi\omega_1|d_{12}|^{\,2}n}{\hbar c\gamma_1} ,\
\alpha_2=\frac{2\pi\omega_2|d_{23}|^{ 2}n}{\hbar c\gamma_2}
\end{equation}

We undertake the mean field limit for both the fields $i.e.$ $\alpha_i L\rightarrow 0$, $T_{i}\rightarrow 0$ and $\delta_{i}\rightarrow 0$, which essentially refer to spatially uniform fields that change little in each pass through the cavity, however, owing to the large cavity photon lifetime the photons undertake many passes inside the cavity~\cite{bonifacio1976} resulting in significant cooperative effects. The governing equations of motion involving the scaled input and output fields are 

\begin{eqnarray}
\frac{\partial x_{1}}{\partial  t'}& = & \kappa_{1}\left[-x_{1}(1+i \theta_{1})+y_{1} + 2 i C_{1} \rho_{12}\right] ~\label{finaleqn1},\\
\frac{\partial x_{2}}{\partial  t'}& = & \kappa_{2}\left[-x_{2}(1+i \theta_{2})+y_{2} +  2 i C_{2} \rho_{23}\right] ~\label{finaleqn2}.
\end{eqnarray}

where the dimensionless input/output fields are defined as 

\begin{eqnarray}
\nonumber
x_1 &=& \frac{d_{12}.E_1^{out}}{\hbar\gamma_1\sqrt{T_1}}, \    x_2=\frac{d_{23}.E_2^{out}}{\hbar\gamma_2\sqrt{T_2}}, \\
y_1 &=& \frac{d_{12}.E_1^{in}}{\hbar \gamma_1\sqrt{T_1}}, \   y_2=\frac{d_{23}.E^{in}_2}{\hbar\gamma_2\sqrt{T_2}}.
\end{eqnarray}

\begin{figure}[thb]
\centering
\includegraphics[width=0.45\textwidth]{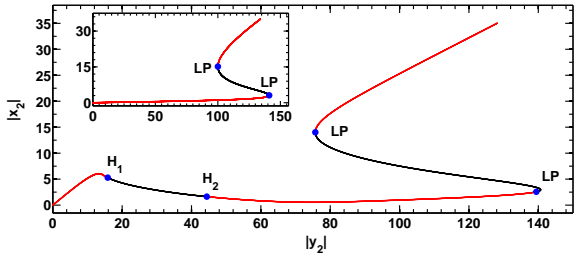}
\caption{(Color online) The bifurcation points associated with the bistable response for the field at $\omega_2$ while the input field at $\omega_1$ is held constant. The parameters are $C_1=200,C_2=200,|y_1|=23$. Inset: The bistable response in absence of feedback to field at $\omega_1$ ($C_1=0,C_2=200$) with $|y_1|=1.5$ and other parameters are $\gamma_i=1,\kappa_i=1,\Delta_1=0,\theta_1=0,\Delta_2=4,\theta_2=-3$ where $i=1,2$.}
\label{wwoffig}
\end{figure}

The last terms in Eqs.~(\ref{finaleqn1},\ref{finaleqn2}) contain the atom-cavity coupling via the cooperative parameters  $C_1=\alpha_1L/2T_1$ and $C_2=\alpha_2L/2T_2$ along the two transitions. The atom-field interaction is governed by the density matrix equations of motion 

\begin{equation}
\frac{\partial\rho}{\partial t'} =-\frac{i}{\hbar} \left[{\hat H},\rho \right]+\cal{\hat L}\rho
~\label{density}
\end{equation}

where the total Hamiltonian $\hat H$ is

\begin{eqnarray}  
\hat H&=&\hat H_{at}+\hat H_{int},\nonumber\\
\hat H_{at}&=&\hbar\omega_{13}|1\rangle\langle1|+\hbar\omega_{23}|2\rangle\langle2|,\\
\hat H_{int}&=&-\vec d \cdot \vec E = -\hbar G_1|1\rangle\langle2|-\hbar G_2|2\rangle\langle3|+h.c.\nonumber
\end{eqnarray}

The terms $\hbar \omega_{13}(\hbar \omega_{23})$ correspond to the energy levels of bare atom measured from the ground state $|3\rangle$, and the interaction Hamiltonian $\hat H_{int}$ is given in the dipole approximation, involving the Rabi frequencies

\begin{equation}
\nonumber
G_1=\frac{d_{12}.E_1}{\hbar} ,\ G_2=\frac{d_{23}.E_2}{\hbar}.
\end{equation}

The relaxation processes like spontaneous emission and dephasing of the atomic coherence are contained in the Liouville operator $\cal{\hat L}$ of Eq.~(\ref{density}).We explicitly enumerate the equations of motion of density matrix elements under the rotating wave approximation  

\begin{eqnarray}
\frac{\partial\rho_{11}}{\partial t'} &=&-2\gamma_1\rho_{11}+i G_{1} \rho_{21}-i G_1^* \rho_{12} \nonumber\\
\frac{\partial\rho_{12}}{\partial t'} &=&-(\gamma_1+\gamma_2+i \Delta_1)\rho_{12}+i G_{1}(\rho_{22}-\rho_{11})\nonumber\\
& &-i G_2^* \rho_{13} \nonumber\\
\frac{\partial\rho_{13}}{\partial t'} &=&-\left[\gamma_1+i(\Delta_1+\Delta_2)\right]\rho_{13}+i G_{1} \rho_{23}\nonumber\\
& &-i G_2 \rho_{12}\nonumber\\
\frac{\partial\rho_{22}}{\partial t'} &=&2 \gamma_1 \tilde\rho_{11}-2 \gamma_2 \rho_{22}-i G_1 \rho_{21}+i G_1^* \rho_{12}\nonumber\\
& &+i G_2\rho_{32}-i G_2^* \rho_{23} \nonumber\\
\frac{\partial\rho_{23}}{\partial t'} &=&-(\gamma_2+i\Delta_2)\rho_{23}+i G_2(\rho_{33}-\rho_{22})\nonumber\\
& &+i G_1^* \rho_{13} \nonumber\\
\frac{\partial\rho_{33}}{\partial t'} &=&2 \gamma_2 \rho_{22}-i G_2\rho_{32}+i G_2^* \rho_{23} 
\label{diffeqn}
\end{eqnarray}

where the atomic detunings are given by $\Delta_1=\omega_{12}-\omega_1$ and $\Delta_2=\omega_{23}-\omega_2$, and 
2$\gamma_1$ and $2\gamma_2$ are the spontaneous emission rates from the levels $|1\rangle$ to $|2\rangle$ and $|2\rangle$ to $|3\rangle$,  respectively, thus considering only the radiative relaxation processes in the dilute atomic gas active medium. The schematic of the level scheme as well as the the two cavities are given in the companion paper~\cite{aswath_1}. All the frequency units are normalized with respect to the atomic decay $\gamma_2$, unless specified otherwise. Furthermore, we undertake a detailed non-linear dynamical analysis of the system described above.

\section{Numerical modelling and stability map}
\label{secnume}

In order to understand the nonlinear  dynamical aspects of this system we solve the field equations Eqs~(\ref{finaleqn1},\ref{finaleqn2}) self consistently with the atomic evolution given in Eqs.~(\ref{diffeqn}). Investigation of the non-linear dynamics is carried out by computing the response (such as ouput fields $x_i$) for a  given set of input fields $y_1$ and $y_2$, for a certain choice of the atomic parameters such as $\Delta_i$ and $\gamma_i$, the cavity parameters $\theta_i$, and $\kappa_i$ and the atom-cavity coupling governed by the cooperative parameters $C_i$. The Newton-Raphson method is used to obtain the solutions of the coupled nonlinear system of equations in the steady state limit. The atom-field cavity interaction dictates the phases and amplitudes of both the 
output fields. The multiplicity of the underlying solutions needs a careful handling with regard to its numerical computation and is described in detail in the companion paper~\cite{aswath_1}. In essence, this system does not permit an apriori choice of the output fields, which is undertaken conventionally to compute the corresponding input field (in systems involving feedback for one field).

\begin{figure}[thb]
\begin{minipage}[b]{0.45 \textwidth}
\includegraphics[width=\textwidth]{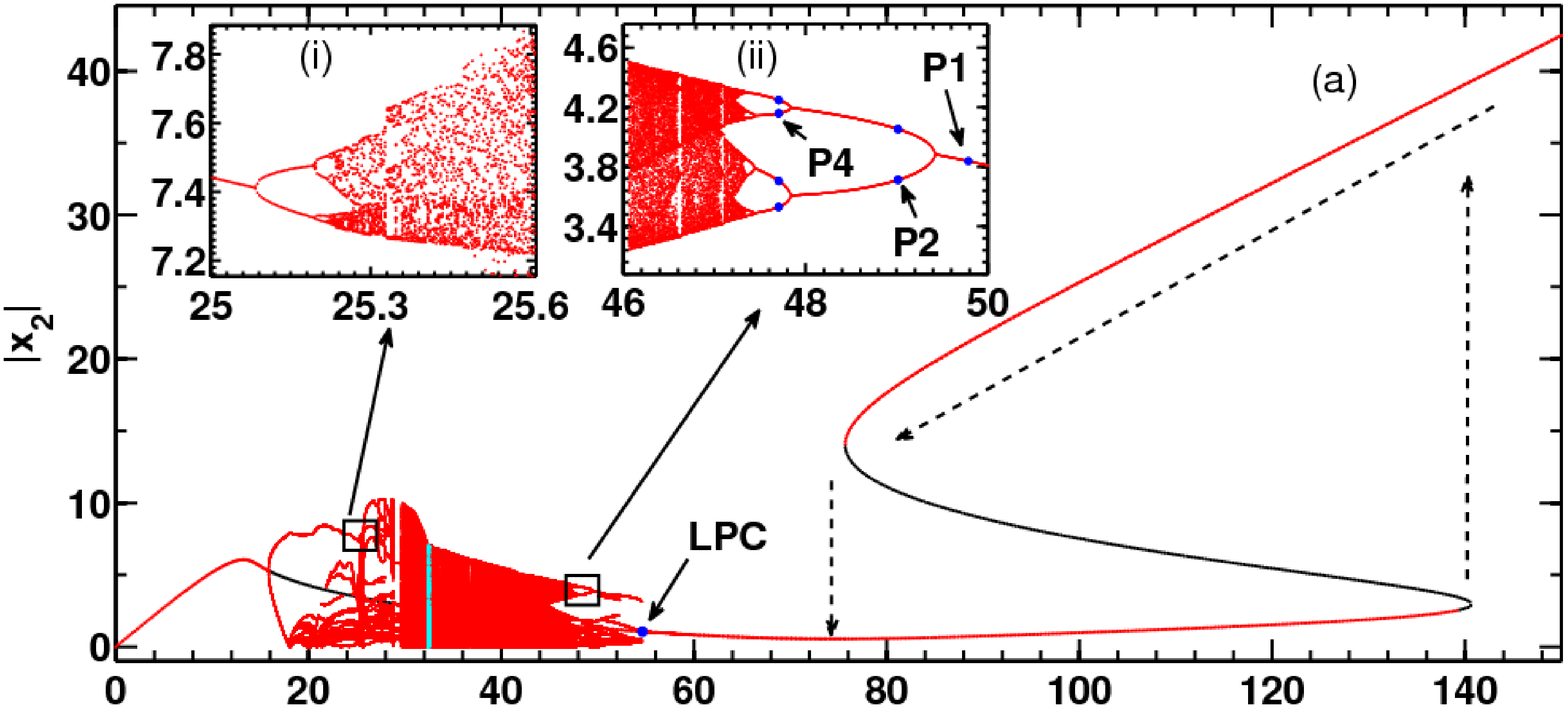}
\end{minipage}
\begin{minipage}[b]{0.45 \textwidth}
\includegraphics[width=\textwidth]{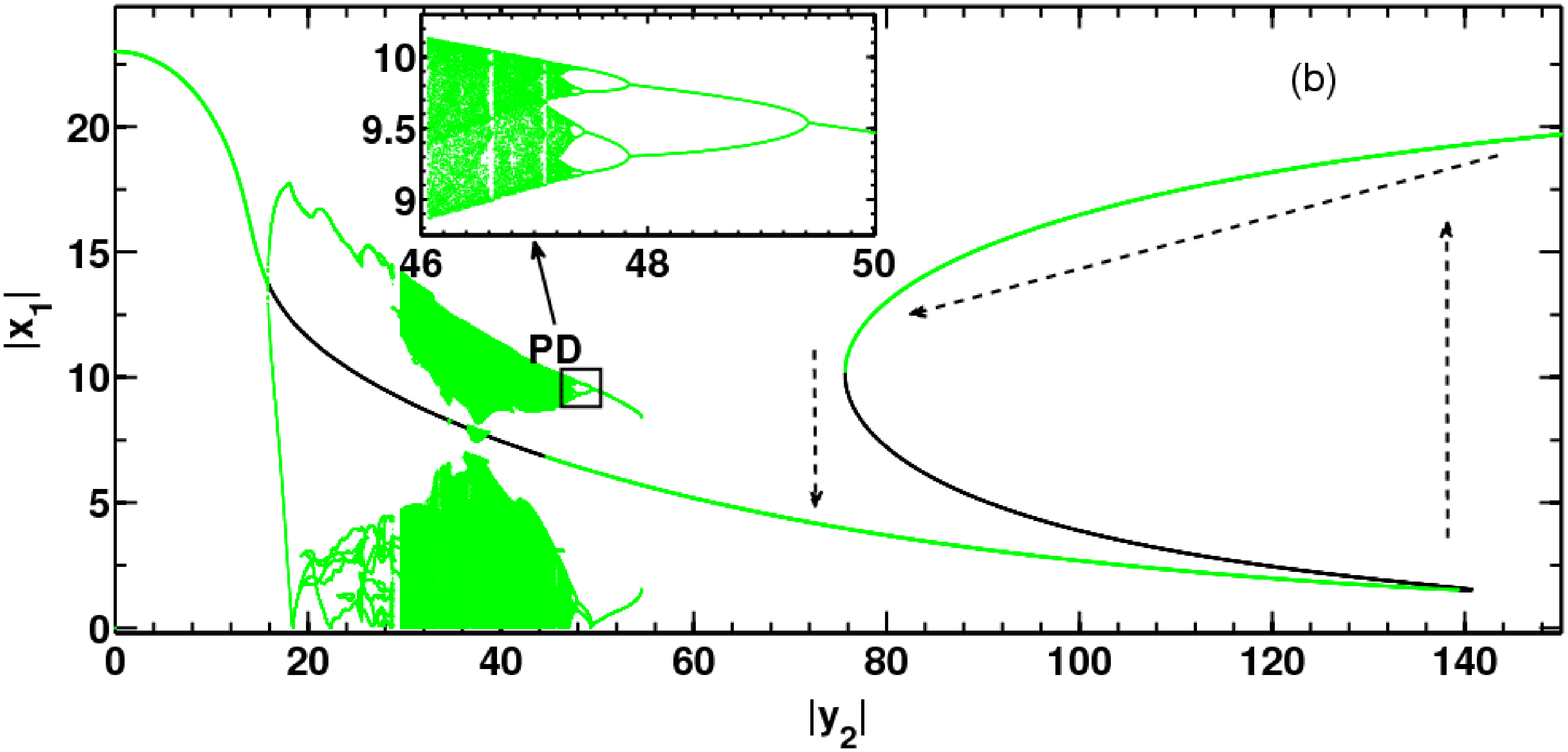}
\end{minipage}
\caption{(Color online)Bifurcation diagrams of the cavity output fields. (a) $|x_2|$ versus $|y_2|$ and (b) $|x_1|$ versus $|y_2|$, for the same parameters as in  Fig.~\ref{wwoffig}}
\label{lbifurfig} 
\end{figure}

The computations were undertaken using two independent numerical tools, one involving Fortran libraries associated with EISPACK~\cite{eispack} and the other using MATLAB. Furthermore, we have also used the MATLAB  continuation package MATCONT~\cite{mat1} for study of the  nonlinear dynamical  aspects of this system. We identify a series of fixed points that exhibit interesting bifurcations such as Hopf points, Limit points as shown in Fig.~\ref{wwoffig}.
 
\begin{figure*}[thb]
\includegraphics[width=1.\textwidth]{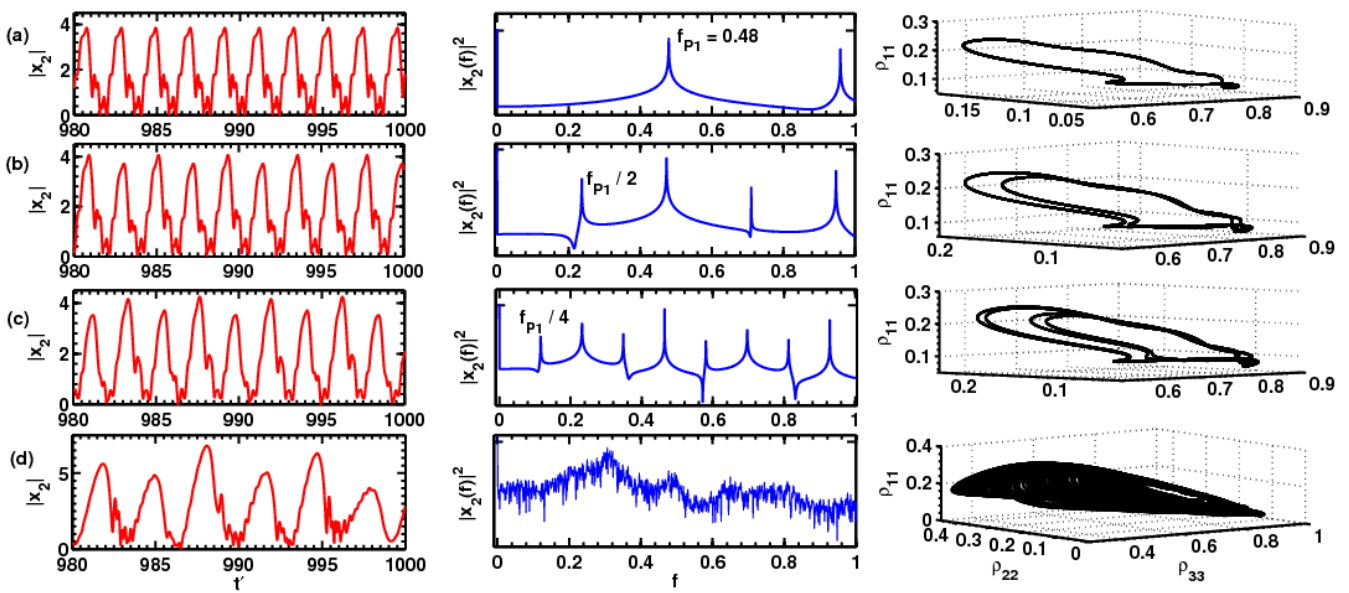}
\caption{(Color onlone)Time evolution of output field $x_2$ at $\omega_2$ are shown in the first column, the corresponding spectrum (in arbitrary units) and the population phase plots ($\rho_{11}$ versus $\rho_{22}$ and $\rho_{33}$) are shown in the middle and the last column respectively. Plots in each row corresponds to various values of $|y_2|$ involved in the sequence of period doubling route ultimately leading to chaos, (a) Period one with $|y_2|=49.78$, (b) Period two with $|y_2|=49.01$, (c) Period four with $|y_2|=47.70$, and (d) Chaos with $|y_2|=32.47$. The other parameters are  same as those for  Fig.~\ref{wwoffig}}
\label{dynlow} 
\end{figure*}

We present the highlights of the linear stability analysis in the form of a detailed stability map shown in Fig.~\ref{domain-map}. The unstable regions and its neighborhood are quite often related to a variety of dynamical behaviour such as periodic, quasi-periodic or even chaotic dynamics, and hence a detailed bifurcation analysis is undertaken. Specifically, the periodic self-pulsing dynamics is further characterized by the Floquet multipliers that distinguish between the stable and unstable limit cycles, and the Lyapunov exponents are calculated in order to identify and characterize chaotic behavior~\cite{strogatz}. All these nonlinear dynamical features are independently confirmed by integrating the time-dependent equations Eqs.~(\ref{finaleqn1}),(~\ref{finaleqn2}),  and (~\ref{diffeqn})
using the fourth order Runge-Kutta technique~\cite{Numericalrecipe}.
                                                                                                                     
We believe the stability domain map would facilitate the experimental realization of these diverse nonlinear dynamical features. The parameter space associated with the system is exceedingly large owing to twelve different physical parameters $\gamma_{1,2}$, $\Delta_{1,2}$, $\theta_{1,2}$, $\kappa_{1,2}$, $C_{1,2}$,  $E^{in}_{1,2}$ each of which can be varied independently. The stability map in Fig.~\ref{domain-map} indicates the range of just two of 
these parameters $\Delta_2$  and  $\theta_2$ for which the system exhibits stable switching (blue region), periodic or quasi-periodic (green region) and chaotic (red region). One can clearly see the islands (green) of stable self-pulsing within the chaotic region (red). The dashed black line indicates the dynamics that we further explore 
and is discussed below in detail along with the associated bifurcation diagram. The stability domain map allows one to pickup the appropriate parameter regime that corresponds to a desired dynamical behavior, such as self-pulsing, stable switching or chaotic output, even a complete avoidance of chaotic region is possible by a judicious choice of the trajectory in the multidimensional parameter space.

\section{Non-linear dynamics} 
\label{secrana}

In order to understand the route to chaos we consider the bifurcation diagram along the dashed line in Fig.~\ref{domain-map}, which is shown in Fig.~\ref{undfigroute}(a). As one varies the detuning of the field $\Delta_2$ while maintaining a constant cavity detuning, we obtain a period doubling cascade for the values of $\Delta_2$ within $[0:1.2]$ and inverse periodic doubling around $\Delta_2 \approx 8.5$. The largest among the Lyapunov exponents corresponding to the regions of self-pulsing are close to zero. The chaotic domains are clearly identified by the positive largest Lyapunov exponents as shown in  Fig.~\ref{undfigroute}(b). The self-pulsing and chaotic dynamics occur  intermittently and the  windows of stable periodic  self-pulsing within a largely chaotic region are clearly seen in Fig.~\ref{undfigroute}(b). Beyond $\Delta_2 \approx 8.5$ all the Lyapunov coefficients become negative indicating that the system quickly moves towards a stable fixed point behavior independent of the initial conditions, in our case it corresponds to stable switching.
 
\begin{table}
\begin{tabular}{|c|c|c|c|}
\hline 
%$|y_2|$  &  Floquet Multipliers &  Floquet Multipliers &  Floquet Multipliers\\
 $|y_2|$  & Period-one & Period-two & Period-four\\
\hline \hline
$49.78$ &  $0.98 + 0 i$ & -  & - \\
\hline
$49.01$ &  $-1.46 + 0 i$  & $0.99 + 0 i$  & - \\
\hline
$47.70$ &  $-2.58 + 0 i$ & $-1.77 + 0 i$ & $0.87 + 0 i$ \\
\hline
\end{tabular}
\caption{ The Floquet multipliers for the first few points of periodic doubling cascade for the input field indicated in the inset.(ii) of Fig.~\ref{lbifurfig}(a)}
\label{table1}
\end{table}

We illustrate the essential nonlinear dynamical features where the magnitude of the input field $y_1 = 23$, while the other incident field $y_2$ is varied. In the S-shaped bistable response we obtain a  hump like feature, and  the associated linear stability analysis is shown in Fig.~\ref{wwoffig}, black (red) color indicate unstable (stable) fixed points. The corresponding bistable curve without the cavity feedback for the field coupling upper transition (systems with cooperative parameter $C_1 = 0$) is shown in the inset of Fig.~\ref{wwoffig}. This hump like feature under appropriate parameter regime transforms into negative hysteretic bistable response. Furthermore, this regime is also associated with non-linear dynamical behavior, and is intimately related to the cavity assisted 
inversion as discussed in the companion paper~\cite{aswath_1}. The hump exhibits Hopf bifurcations and these effects are absent in OB models involving feedback for one field(inset of Fig.~\ref{wwoffig}), however we note that, this 
feature is a result of an intricate interplay of phases of both the fields. If one demands that both the output fields are real, as in Ref.~\cite{xiao-joshi}, this hump like feature as well as the nonlinear dynamical features disappear. We  indicate the bifurcation points such as the Hopf points(H),  limit points(LP) in the figure. In the unstable regime corresponds  lower cooperative branch, one obtains self pulsing  as well as chaos at low input light levels. The bifurcation diagram associated with Fig.~\ref{wwoffig} is shown in Fig.~\ref{lbifurfig}. We also expand two illustrative regions of the bifurcation diagram that indicate the onset of chaos. With increasing input field $y_2$ the system loses stability of stable fixed point solutions at first Hopf point ($H_1$) and exhibits self-pulsing behavior. This is indicative of a supercritical Hopf bifurcation at $H_1$. The Hopf points indicate the onset of periodic behavior, and in between the Hopf points we observe a period doubling route to chaos. The stability of the periodic behavior is established using Floquet multipliers, and  chaos is confirmed using Lyapunov Exponents as well as the output field spectrum. 
 
There is a simultaneous existence of stable fixed point solutions as well as self-pulsing limit cycles beyond the second Hopf point $H_2$. For certain range of $y_2$ magnitude within $(48:55)$ depending on the initial condition the system can be driven to exhibit either periodic self-pulsing or regular stable switching. A transition from periodic behavior to fixed point solution behavior occurs as a Saddle-node/fold bifurcation takes place beyond $H_2$ at $y_2=54.6$, and is represented as limit point of cycles (LPC) (see Fig.~\ref{lbifurfig}). At LPC two limit cycles (one stable and one unstable) coalesce and annihilate each other, leading to pure fixed point solutions~\cite{strogatz}. The stability of limit cycles indicated by the Floquet multipliers remaining with in unit circle and transition out of unit circle implies unstable limit cycles. The Floquet multipliers associated with the stable and unstable limit cycles (at $|y_2|=51$ before coalesce at LPC) are $0.3639 + 0 i$ and $-1.7997 + 0 i$ respectively.

The onset of chaos indicated in Fig.~\ref{lbifurfig} occurs as one  decreases the input field $y_2$ magnitude from 50 (see inset (ii)). At first one obtains self pulsing output of period 1, which transforms into a self pulsing output of period 2 and then to period 4, as shown in Fig.~\ref{dynlow} (a-c). The genesis of new frequencies in the spectrum of the output field (second column) at $f_{P2}=f_{P1}/2$,  $f_{P4}=f_{P1}/4$ and its multiples, apart from the dominant frequency $f_{P1}=0.48$, clearly indicates a period doubling route. The dashed arrows in Fig.~\ref{lbifurfig} indicate conventional bistable switching. As both the fields switch simultaneously, one can utilize this system for controllable switching at two distinct optical frequencies in optical communication applications. The period doubling route to chaos is further substantiated by the three dimensional phase plots (third column) involving the population of the three atomic states which show closed curves involving one, two and four loops, respectively. As we further decrease the input field $y_2$ magnitude to $32.47$ (indicated by thick line with cyan color in Fig.~\ref{lbifurfig}(a)) one observes chaos and the corresponding two largest Lyapunov exponents of the system are $0.3356$ and $0.0031$. The spectrum of the output field becomes continuous  as well as the population phase plot gets filled up clearly establishing the existence of chaos. This is further corroborated by the loss of stability of periodic behavior (limit cycles)~\cite{kuznetsov} as shown in Table.~\ref{table1}, wherein the associated Floquet multipliers cross out of the unit circle along the negative real axis. The Floquet multipliers having zero imaginary part and transition out of the unit circle along negative imaginary axis precludes the existence of quasi periodicity, thus establishing a periodic-doubling route to chaos as indicated.The nonlinear dynamics of both the fields closely resembles each other, and one obtains periodic self-pulsing to chaotic dynamics for both the fields 
$x_1$ and  $x_2$ simultaneously. 

\begin{figure}[thb]
\includegraphics[width=0.5\textwidth]{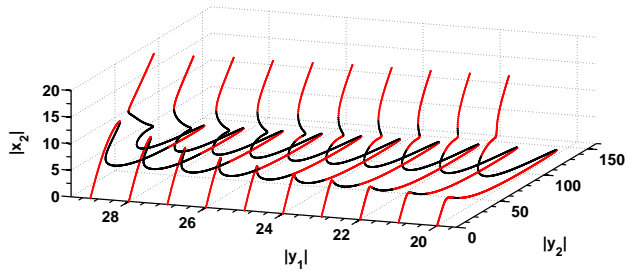}
\caption{(Color online) The development of negative hysteresis and multistability for the field at $\omega_2$ with $\Delta_1=4,\Delta_2=0$. Each curve corresponds to a particular value of the input field at $\omega_1$ whose magnitude is varied from $20$ to $30$, and the other parameters are same as in Fig.~\ref{wwoffig}} 
\label{multi} 
\end{figure}

This system offers a variety of control mechanisms which allows one to access any desired dynamics by merely changing a few pertinent parameters. The inclusion of finite atomic detuning $\Delta_1$ could even lead to multistability as seen in Fig.~\ref{multi}. The figure also indicates the various domains of stability where interesting nonlinear dynamical behavior occurs. A detailed analysis of the multistable behavior will be presented elsewhere.

\begin{figure}[thb]
\centering
\includegraphics[width=0.5\textwidth]{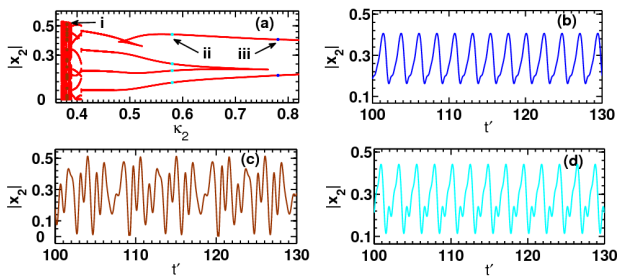}
\caption{(Color online)(a)The bifurcation diagram for the field at $\omega_2$ as the $\kappa_2$ cavity decay is tuned. (b),(c) and (d) shows the time evolution of output field magnitude for $\kappa_2=0.7808,0.3008,0.5803$ (marked as $i, iii$ and $ii$ in (a)) for $C_1=1000,C_2=100,|y_1|=25,|y_2|=13,\theta_2=0,\Delta_2=0$ and the other parameters are same as in Fig.~\ref{domain-map}.}
\label{bifurk2}
\end{figure}

\begin{figure}[thb]
\includegraphics[width=0.5\textwidth]{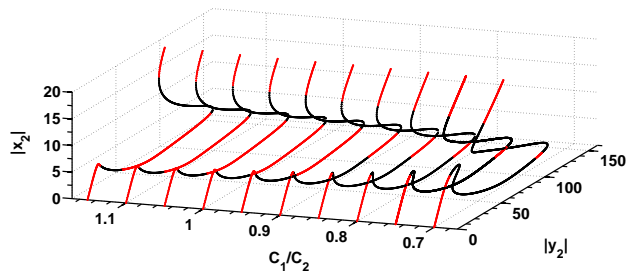}
\caption{(Color online) The hump like feature progressively transforms into a negative hysteresis for comparable values of $C_1$ and $C_2$, and other parameters are same as Fig.~\ref{wwoffig}}
\label{effofcop2} 
\end{figure}

We indicate other ways to control the stability as well as nonlinear dynamical aspects in this system. One can obtain negative hysteretic behavior,  with the associated non-linear dynamical features in its neighborhood. One can obtain all the above mentioned non-linear dynamical features by varying the cavity decays $\kappa_1$ and/or $\kappa_2$. Such control over the nonlinear dynamical response is illustrated in the bifurcation diagram shown in Fig.~\ref{bifurk2} which involves $\kappa_2$ as the control parameter which can be used to switch {\em on} or {\em off} the chaotic dynamics. It should be noted that including finite cavity detuning ($\theta_2$) aids in obtaining all the  non linear dynamical features with comparable values of the co-operative parameters $C_1$ and $C_2$ as shown in Fig.~\ref{effofcop2}. 

\section{Conclusions and remarks}
\label{secconc}

We have demonstrated a new regime of nonlinear dynamical response at low input light levels  for two-photon  double cavity optical bistability with three-level ladder atomic system as the active medium. Independent feedback is applied for both the fields interacting with the atom. We present bifurcation diagrams that allows one to access the desired regime of dynamics for both the cavity fields. We prove the period-doubling route to chaos in this new regime associated with the lower cooperative branch. A control paradigm based on careful maneuvering of parameters so as to traverse across phase space in order to obtain {\em any} desired dynamics is demonstrated. The system exhibits negative as well as positive hysteresis bistable response, stable periodic self-pulsing and  chaotic dynamics, apart from the conventional normal switching and multistability.

\section{acknowledgments}

We gratefully acknowledge the help provided by Drs. Supriyo Pal, Pankaj Wahi and M.K. Verma, 
particularly with regard to the
Non-linear dynamical aspects of this work.

\end{document}